\documentclass[conference]{IEEEtran}

\usepackage[noadjust]{cite}
\ifCLASSINFOpdf
   \usepackage[pdftex]{graphicx}
\else
\fi
\usepackage[cmex10]{amsmath}
\usepackage{array}
\usepackage{mdwmath}
\usepackage{mdwtab}
\usepackage{eqparbox}
\usepackage{subfigure}
\usepackage{caption}
\usepackage{url}
\usepackage{stackrel}
\usepackage{etoolbox}
\usepackage{enumerate}
\usepackage{amssymb}
\usepackage{tikz, pgfplots}
\usetikzlibrary{calc}
\usetikzlibrary{arrows}
\usetikzlibrary{arrows,positioning} 
\usetikzlibrary{decorations.markings}
\usetikzlibrary{shapes}
\tikzset{
    >=stealth',
    punkt/.style={
           rectangle,
           rounded corners,
           draw=black, very thick,
           text width=6.5em,
           minimum height=2em,
           text centered},
    pil/.style={
           ->,
           thick,
           shorten <=2pt,
           shorten >=2pt,}
    pir/.style={
           <-,
           thick,
           shorten <=2pt,
           shorten >=2pt,}
}
\definecolor{KeynoteRed}{rgb}{.678,.051, .051}
\definecolor{KeynoteBlue}{rgb}{0.008, 0.443, 0.60}
\definecolor{KeynoteLightblue}{rgb}{.635, .914, .973}
\definecolor{KeynoteYellow}{rgb}{0.859, 0.584, 0.212}
\definecolor{KeynoteYellow}{rgb}{0.859, 0.584, 0.212}
\definecolor{KeynoteSlate}{rgb}{0.239, 0.271, 0.322}
\definecolor{KeynoteGray}{rgb}{0.498, 0.529, 0.529}
\definecolor{KeynoteGreen}{rgb}{0.18, 0.5, 0.08}
\definecolor{KeynoteTextGray}{rgb}{0.325, 0.325, 0.325}
\definecolor{KeynoteLightGray}{rgb}{0.706, 0.706, 0.706}
\definecolor{KeynoteBlueGray}{rgb}{0.471, 0.533, 0.620}
\definecolor{ECEpurple}{rgb}{.169, .18, .455}
\definecolor{ECEcyan}{rgb}{.41, .62, .72}
\definecolor{ECEgray}{rgb}{.788, .827, .859}
\definecolor{ECEblueGray}{rgb}{61.2, 70.6, 70.6}
\definecolor{ECEblueGray}{rgb}{61.2, 70.6, 70.6}
\definecolor{RiceBlue}{rgb}{0, .14, .41}

\newcommand{\LenOneTx}{{L}_{T_1}}
\newcommand{\LenTwoTx}{{L}_{T_2}}
\newcommand{\LenTx}{{L}_{T_j}}
\newcommand{\LenOneRx}{{L}_{R_1}}
\newcommand{\LenTwoRx}{{L}_{R_2}}
\newcommand{\LenBS}{{L}_\mathsf{BS}}
\newcommand{\LenU}{{L}_\mathsf{Usr}}
\newcommand{\LenRx}{{L}_{R_i}}

\newcommand{\LenTwoTxPrime}{{L'}_{T_2}}

\newcommand{\LenOneRxPrime}{{L'}_{R_1}}

\newcommand{\ClusterR}[1]{\Omega_{R_{#1}}}

\newcommand{\ElevTx}{\Theta_{T_{ij}}}
\newcommand{\ElevRx}{\Theta_{R_{ij}}}
\newcommand{\ElevT}[1]{\Theta_{T_{#1}}}
\newcommand{\ElevR}[1]{\Theta_{R_{#1}}}

\newcommand{\IntT}[1]{\Psi_{T_{#1}}}
\newcommand{\IntR}[1]{\Psi_{R_{#1}}}
\newcommand{\IntTPrime}[1]{\Psi'_{T_{#1}}}
\newcommand{\IntRPrime}[1]{\Psi'_{R_{#1}}}
\newcommand{\IntTx}{\Psi_{T_{ij}}}
\newcommand{\IntRx}{\Psi_{R_{ij}}}
\newcommand{\IntFwd}{\Psi_\mathsf{Fwd}}
\newcommand{\IntBack}{\Psi_\mathsf{Back}}
\newcommand{\TxSpace}[1]{\mathcal{T}_{#1}}
\newcommand{\RxSpace}[1]{\mathcal{R}_{#1}}

\newcommand{\ScatOp}[1]{\mathsf{H}_{#1}}

\newcommand{\Scat}[1]{H_{#1}}
\newcommand{\TxTwoPreSpace}{\mathcal{P}_{12}}

\newcommand{\FlowOne}{\mathsf{Flow}_1}
\newcommand{\FlowTwo}{\mathsf{Flow}_2}
\newcommand{\AngleR}[1]{\theta_{R_{#1}}}
\newcommand{\AngleT}[1]{\theta_{T_{#1}}}

\newcommand{\preim}[2]{{#2}^{\leftarrow}(#1)}

\newcommand{\dOneMax}{d_1^{\sf max}}
\newcommand{\dTwoMax}{d_2^{\sf max}}
\newcommand{\dSumMax}{d_{\sf sum}^{\sf max}}
\newcommand{\dSum}{d_{\sf FD}^{\sf sum}}
\newcommand{\D}{\mathcal{D}_{\sf FD}}
\newcommand{\DHD}{\mathcal{D}_{\sf HD}}

\newcounter{MYtempeqncnt}

\newtheorem{rem}{Remark}

\newtheorem{thm}{Theorem}

\newtheorem{lem}{Lemma}

\newtoggle{ZChanModel}
\begin{document}

\title{A Signal-Space Analysis of Spatial Self-Interference Isolation for Full-Duplex Wireless\vspace{0pt}}

%\author{Evan~Everett and Ashutosh~Sabharwal}

\author{\IEEEauthorblockN{Evan Everett}
\IEEEauthorblockA{Rice University}
\and
\IEEEauthorblockN{Ashutosh~Sabharwal}
\IEEEauthorblockA{Rice University}

}

\toggletrue{ZChanModel}

\maketitle

\begin{abstract}
%%%\boldmath
The challenge to in-band full-duplex wireless communication is managing self-interference. %: a node's transmit signal appears at its own receiver as high-powered interference.  
Many designs have employed spatial isolation mechanisms, such as shielding or multi-antenna beamforming, to isolate the self-interference wave from the receiver. Such spatial isolation methods are effective, but by confining the transmit and receive signals to a subset of the available space, the full spatial resources of the channel be under-utilized, expending a cost that may nullify the net benefit of operating in full-duplex mode. In this paper we leverage an antenna-theory-based channel model to analyze the spatial degrees~of~freedom available to a full-duplex capable base station, and observe that whether or not spatial isolation out-performs time-division (i.e. half-duplex) depends heavily on the geometric distribution of scatterers. Unless the angular spread of the objects that scatter to the intended users is overlapped by the spread of objects that backscatter to the base station, then spatial isolation outperforms time division, otherwise time division may be optimal. 

%We observe that the degree-of-freedom is highly dependent on the extent of overlap between 
%We derive the degree-of-freedom region for the scenario in which two users, one uplink and one downlink, are served by the full-duplex base~station, and observe that full-duplex spatial isolation can indeed 

\end{abstract}
% IEEEtran.cls defaults to using nonbold math in the Abstract.

% Note that keywords are not normally used for peerreview papers.
%\begin{IEEEkeywords}
%Full-duplex, coherence bandwidth, self-interference.
%\end{IEEEkeywords}

\section{Introduction}
% !TEX root = ISIT_04_main.tex

\begin{figure}[htbp]
\begin{center}
\resizebox{0.5\textwidth}{!}{
	\input{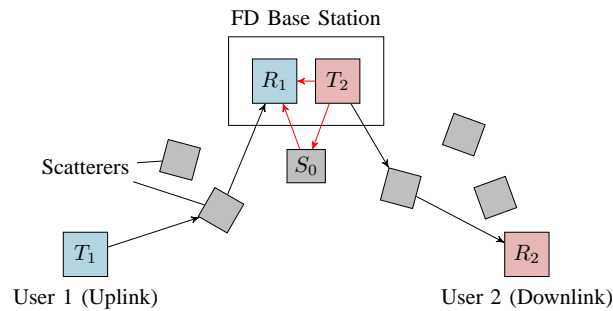}
}
\caption{Three-node full-duplex model}
\label{fig:threeNode}
\end{center}
\end{figure}

Consider the communication scenario depicted in Figure~\ref{fig:threeNode}. 
User 1 wishes to transmit uplink data to a base station, and User 2 wishes to receive downlink data from the same base station. 
%If the base station is half-duplex, then it must either service the users in orthogonal time slots or in orthogonal frequency bands. 
If the base station can operate in full-duplex mode, i.e., transmits and receives at the same time in the same band, then it can enhance spectral efficiency by servicing both users simultaneously. % We expect this ``three-node full-duplex'' scenario to be among the first commercial applications of full-duplex, as the burden of full-duplex is placed only on the infrastructure base station and space-constrained mobile user devices can remain half-duplex.% while still benefitting from full-duplex operation at the base station. 
To cancel the high-powered self-interference, the knowledge of the transmit signal can be used to perform self-interference cancellation. 
%The challenge presented in full-duplex wireless communication is high-powered \emph{self-interference}. Since a node has knowledge of its own transmit signal, self-interference \emph{cancellation} can be performed. 
However, experimental studies have shown that cancellation alone is often insufficient to realize the ideal doubling of capacity  over half-duplex \cite{Duarte11FullDuplex,SahaiPhaseNoiseDraft}. Thus methods to create spatial isolation between transmit and receive antennas, like multi-antenna beamforming \cite{Bliss07SimultTX_RX,Day12FDMIMO}, directional antennas \cite{Everett11Asilomar}, and shielding via absorptive materials \cite{Everett2013PassiveSuppressionFD}, are also employed. Unlike cancellation, spatial isolation may consume channel resources that could have otherwise been leveraged for signal-of-interest communication. 

%\evanNote{In wireless communication, electromagnetic scattering off of physical objects provides multiple propagation paths from a transmitter to a receiver, which can be leveraged to increase capacity via spatial multiplexing.}
Consider the example illustrated in Figure~\ref{fig:threeNode}. 
%The environment contains numerous scatterers which allow nodes with multiple antennas to increase capacity via spatial multiplexing. 
The direct path from the base station transmitter, $T_2$, to its receiver $R_1$, can be suppressed by creating a radiation pattern with a null in the direction of $R_1$, but there will also be self-interference due to reflections from the scatterers. The self-interference caused by scatterer $S_0$ in Figure~\ref{fig:threeNode} could be avoided by creating a null in the direction of $S_0$. However losing access to that scatterer could lead to a less rich scattering environment, diminishing the spatial degrees of freedom of the uplink or downlink. 
%Moreover, creating the null consumers antenna resources at the base station that could have been leveraged for spatial multiplexing, also diminishing the spatial degrees of freedom of the uplink or downlink. This example us leads to pose the following question

\textbf{Question}: 
Under what scattering conditions can spatial isolation be leveraged to provide a degree-of-freedom gain over half duplex?
%To answer the above question, we formulate the problem as follows.
More specifically, given a constraint on the \emph{area} of the arrays at the base station and  at the User 1 and User 2 devices, and given a characterization of the \emph{spatial distribution} of the scatterers in the environment, what is the uplink/downlink degree-of-freedom region when the only self-interference mitigation strategy is spatial isolation?% (i.e. inner bound on DoF with cancellation)?

\textbf{Modeling Approach:}
%The conventional statistical MIMO model of a discrete array with a random channel matrix, is not well suited for answering the above question. \evanNote{I need to give two crisp reasons here. I am thinking about this about this.}
To answer the above question we leverage physical channel model developed by Poon, Broderson, and Tse in \cite{TsePoon05DOF_EM,TsePoon06EmagInfoTheory,TsePoon11DOFPolarization}, which we will call the ``PBT'' model.  In the PBT model, instead of constraining the \emph{number} of antennas, a constraint on the \emph{area} of the array is given, and instead of considering a channel matrix drawn from a probability distribution, a channel transfer function which depend on the geometric position of the scatterers relative to the arrays is considered. %In particular, the channel transfer function is contained to obey 
%Therefore the results obtained using the model inform not just the processing at the transmitter and the receiver, but also imforms antenna design and placement. 

\textbf{Contribution}
We extend the PBT model to the three-node full-duplex topology of Figure~\ref{fig:threeNode}, and derive the degree-of-freedom region $\D$, i.e. the set of all achievable uplink/downlink degree-of-freedom tuples. By comparing $\D$ to $\DHD$, the degree-of-freedom region achieved by time-division duplex, we observe that $\DHD\subset\D$ in the following two scenarios:
\begin{enumerate}
\item When the base station arrays are larger than the corresponding user arrays, so that the extra resources used for spatial isolation were not needs for spatial multiplexing, 
\item More interestingly, when the forward scattering intervals and the backscattering intervals are not completely overlapped. In Figure~\ref{fig:threeNode} for example, if there are some directions from which $T_2$'s radiated signal will scatter to the intended receiver, $R_2$, but not backscatter to $R_1$, then $T_2$ can avoid interference by signaling in those directions without having to zero-force to $R_2$.
%
% is a sufficient amount non-overlapping scattering directions. For examples if their are many directions in which $T_2$ can transmit 
\end{enumerate}

%\textbf{Subquestion 2:} Given constrained total array size what portion of the array should be allocated for transmission only, what portion for reception only, and what portion for both? 

%Thus we pose the following general question. 

%\section{Related Work}
%\input{RelatedWork}

\section{System Model}
\label{sec:systemModel}
Here we extend the PBT channel model in \cite{TsePoon05DOF_EM}, which considers a point-to-point topology, to the three-node full-duplex topology of Figure~\ref{fig:threeNode}. 
\subsection{Overview of PBT Model}
The PBT channel model considers a wireless communication link between a transmitter equipped with a unipolarized continuous linear array of length $2L_T$ and a receiver with a similar array of length $2L_R$.
%The scattering environment is characterized by the angle subtended by the scatters. 
The authors observe that there are two key domains: the \emph{array domain}, which describes the current distribution on the arrays, and the \emph{wavevector domain} which describes the field patterns.
Assume the physical objects that scatter the fields radiated from the transmit array to the receive array subtend an angle $\ElevT{}$ at the transmit array an angle $\ElevR{}$ at the receive array. 
%So that 
%so that a wavevector radiated from the transmit array at angle $\theta$ will couple to the receiver only if $\theta \in \ElevT{}$.
%Likewise the objects that scatter the  subtend  angle $\ElevR{}$ at the receive array. 
%
Because a linear array aligned to the $z$-axis array can only resolve the $z$-component, i.e. the $\cos\theta$ component, consider the sets $\Psi_T = \{\cos \theta: \theta\in\ElevT{} \}$ and $\Psi_R = \{\cos \theta: \theta\in\ElevR{} \}$. In  \cite{TsePoon05DOF_EM}, it is shown from the first principles of Maxwell's equations that an array of length $2L_T$ has a resolution of $1/(2L_T)$ over the interval $\Psi_T$, so that the dimension of the transmit signal space of radiated field patterns is $2L_T|\Psi_T|$.
%which are array limited to $2L_T$ and wavevector limited to $\Psi_T$ is $2L_T\Psi_T$. 
Likewise the dimension of the receive signal space is $2L_R|\Psi_R|$, so that the degrees of freedom of the communication link is 
\begin{equation}
d_\mathrm{P2P} = \min\left\{2L_T|\Psi_T|,  2L_R|\Psi_R| \right\}.
\end{equation}

\begin{figure}[htbp]
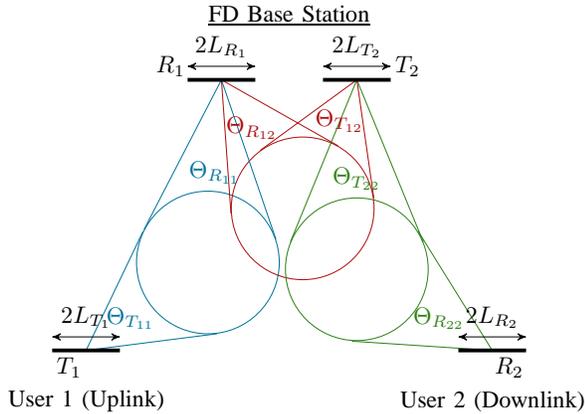

\begin{center}
\scalebox{0.9}{%
\tikzstyle{block} = [draw,fill=KeynoteBlue!30,minimum size=2em]
\tikzstyle{bigBlock} = [draw,fill=KeynoteRed!50,minimum size=1.5em]
\tikzstyle{scatterer} = [draw,fill=gray!50,minimum size=2em]
\tikz[auto, decoration={
  markings,
  mark=at position 1.0 with {\arrow{>}}}
]{
\coordinate (T1) at (-3,0);
\coordinate (R1) at (-1,4);
\coordinate (T2) at (1,4);
\coordinate (R2) at (3,0);
\node [circle,draw,KeynoteBlue] (S11) at (-1.2,1.3) [minimum size=60pt] {};
\node [circle,draw,KeynoteRed] (S12) at (0.2,2.1) [minimum size=60pt] {};
\node [circle,draw,KeynoteGreen] (S22) at (1,1.2) [minimum size=60pt] {};
\draw[KeynoteBlue] (T1) -- (tangent cs:node=S11,point={(T1)},solution=1);
\draw[KeynoteBlue] (T1) -- (tangent cs:node=S11,point={(T1)},solution=2);
\draw[KeynoteBlue] (R1) -- (tangent cs:node=S11,point={(R1)},solution=1);
\draw[KeynoteBlue] (R1) -- (tangent cs:node=S11,point={(R1)},solution=2);
\draw[KeynoteRed] (T2) -- (tangent cs:node=S12,point={(T2)},solution=1);
\draw[KeynoteRed] (T2) -- (tangent cs:node=S12,point={(T2)},solution=2);
\draw[KeynoteRed] (R1) -- (tangent cs:node=S12,point={(R1)},solution=1);
\draw[KeynoteRed] (R1) -- (tangent cs:node=S12,point={(R1)},solution=2);
\draw[KeynoteGreen] (T2) -- (tangent cs:node=S22,point={(T2)},solution=1);
\draw[KeynoteGreen] (T2) -- (tangent cs:node=S22,point={(T2)},solution=2);
\draw[KeynoteGreen] (R2) -- (tangent cs:node=S22,point={(R2)},solution=1);
\draw[KeynoteGreen] (R2) -- (tangent cs:node=S22,point={(R2)},solution=2);
\node[KeynoteBlue] at ($(S11.center)!40pt!(T1)$) {$\ElevT{11}$};
\node[KeynoteBlue] at ($(S11.center)!38pt!(R1)$) {$\ElevR{11}$};
\node[KeynoteRed] at ($(S12.center)!40pt!(T2)$) {$\ElevT{12}$};
\node[KeynoteRed] at ($(S12.center)!40pt!(R1)$) {$\ElevR{12}$};
\node[KeynoteGreen] at ($(S22.center)!38pt!(T2)$) {$\ElevT{22}$};
\node[KeynoteGreen] at ($(S22.center)!40pt!(R2)$) {$\ElevR{22}$};
\path (T1)+(-0.25,-0.25) node {$T_1$}
(T2)+(.75,.20) node {$T_2$}
(R1)+(-.75,.20) node {$R_1$}
(R2)+(+0.25,-0.25) node {$R_2$};
\draw[ultra thick] ($(R1)+(-.5,0)$)--($(R1)+(.5,0)$);
\draw[<->] ($(R1)+(-.5,.20)$)  to node {$2\LenOneRx$} ($(R1)+(.5,.20)$);
\draw[ultra thick] ($(T2)+(-.5,0)$)--($(T2)+(.5,0)$);
\draw[<->] ($(T2)+(-.5,.20)$)  to node {$2\LenTwoTx$} ($(T2)+(.5,.20)$);
\draw[ultra thick] ($(R2)+(-.5,0)$)--($(R2)+(.5,0)$);
\draw[<->] ($(R2)+(-.5,.20)$)  to node {$2\LenTwoRx$} ($(R2)+(.5,.20)$);
\draw[ultra thick] ($(T1)+(-.5,0)$)--($(T1)+(.5,0)$);
\draw[<->] ($(T1)+(-.5,.20)$)  to node {$2\LenOneTx$} ($(T1)+(.5,.20)$);
\node at ($(R1)!0.5!(T2)+(0,.95)$) {\underline{FD Base Station}};
\node at ($(T1)+(0,-0.75)$) {{User 1 (Uplink)}};
\node at ($(R2)+(0,-0.75)$) {{User 2 (Downlink)}};
\iftoggle{ZChanModel}{}{
\node [circle,draw,KeynoteYellow] (S21) at (-0.2,1) [minimum size=50pt] {};
\node[KeynoteYellow] at ($(S21.center)!40pt!(T1)$) {$\ElevT{21}$};
\node[KeynoteYellow] at ($(S21.center)!40pt!(R2)$) {$\ElevR{21}$};
\draw[KeynoteYellow] (T1) -- (tangent cs:node=S21,point={(T1)},solution=1);
\draw[KeynoteYellow] (T1) -- (tangent cs:node=S21,point={(T1)},solution=2);
\draw[KeynoteYellow] (R2) -- (tangent cs:node=S21,point={(R2)},solution=1);
\draw[KeynoteYellow] (R2) -- (tangent cs:node=S21,point={(R2)},solution=2);
}
}
} 
 
\caption{Clustered scattering. Only one cluster for each transmit receive pair is shown to prevent clutter.%\evanNote{if time allows will draw as ellipses rather than circles}
}
\label{fig:clusters}
\end{center}
\end{figure}

\subsection{Extension of PBT Model to Three-Node Full-Duplex}
Figure~\ref{fig:clusters} illustrates our extension of the PBT channel model to the three-node full-duplex topology of Figure~\ref{fig:threeNode}. Let $\FlowOne$ denote the uplink flow from User 1 to the base station, and $\FlowTwo$ denote the downlink flow from the base station to User 2. Let $T_1$ and $R_1$ denote the transmitter and receiver for  $\FlowOne$, respectively, and $T_2$ and $R_2$ denote the the transmitter and receiver for $\FlowTwo$.
Each of the two transmitters $T_j,\ j = 1,2$ is equipped with a linear  array of length $2\LenTx$, and each receiver, $R_i,\ i = 1,2$ is equipped a linear array of length $2\LenRx$.

%At each transmitter and receiver, define a local coordinate system with origin at the midpoint of the array and $z$-axis aligned to the array.
%Let $\AngleT{j} \in [0,\pi]$ denote the elevation angle relative to the $T_j$ array as shown in Figure~\ref{fig:clusters}, and let $\AngleR{i}$ denote the elevation angle relative to the $R_i$ array. Since the linear arrays can only resolve the $\cos\theta$ component of the field, let $t_j \equiv \cos\AngleT{j} \in [-1, 1]$, and likewise $\tau_i \equiv \cos\AngleR{i} \in [-1, 1]$.
\subsubsection{Scattering Intervals}
Let $\AngleT{j} \in [0,\pi]$ denote the elevation angle relative to the $T_j$ and let $\AngleR{i}$ denote the elevation angle relative to the $R_i$ array. %We will see that field pattern radiated from the $T_j$ array will depend on $\AngleT{j}$ only through $\cos\AngleT{j}$. Thus let $t_j \equiv \cos\AngleT{j} \in [-1, 1]$, and likewise $\tau_i \equiv \cos\AngleR{i} \in [-1, 1]$.
As depicted in Figure~\ref{fig:clusters}, $\ElevTx$ denotes the angular spread subtended at transmitter $T_j$ by the physical objects that scatter fields radiated from $T_j$ to $R_i$. %The scattering interval $\ElevTx$ can be thought of as the set of directions that when illuminated by $T_j$ scatters energy to $R_i$. 
Similarly let  $\ElevRx$ denote the corresponding angular spread subtended at $R_i$ by scatterers illuminated by $T_j$.
Thus, we see in Figure~\ref{fig:clusters} that from the point-of-view of the base-station transmitter, $T_2$, $\ElevT{22}$ is the angular interval over which it can radiate signals that will couple to its intended receiver, while $\ElevT{12}$ is the interval in which the radiated signal will bounce back %\ashuNote{If $\ElevT{12}$ is the interval in which the radiated signal will ``bounce back," then what does $\ElevR{12}$ represent?} 
to the base station receiver, $R_1$, as self-interference. %Likewise, from the point-of-view of the base station receiver $\ElevR{11}$ is the interval over which it may receive signals from the User~1 transmitter, $T_1$, while $\ElevR{12}$ is the interval in which self-interference may be present. Clearly, the extent to which the interference intervals and the signal-of-interest intervals overlap will have a major impact on the degrees of freedom of the network. 
We assume that the user devices are hidden from each other such that $\ElevT{21} = \ElevR{21} = \emptyset$. 
In Figure~\ref{fig:clusters}, the six scattering intervals are drawn as being circular and angularly contiguous, but this is purely for the sake of making the figure uncluttered, and need not be the case. 
%
%\ashuNote{This last sentence answers my question that transmit and receive angles may not be same. The reason is that clusters are drawn as circles. Why not draw ellipses - not much harder to depict than circles but allows you to have different transmit/receive angles.}\evanNote{Good point, will do this.}
Because linear arrays can only resolve the cosine of the elevation angle, 
let $t_j \equiv \cos\AngleT{j} \in [-1, 1]$, and likewise $\tau_i \equiv \cos\AngleR{i} \in [-1, 1]$. Denote the ``effective'' scattering interval as 
%$\IntTx \equiv \left\{t\equiv\cos\theta: \theta \in \ElevTx  \right\} \subset [-1,1].$
$$\IntTx \equiv \left\{t_j: \arccos(t_j) \in \ElevTx  \right\} \subset [-1,1].$$
Likewise for the receiver side we denote the effective scattering intervals as
$$\IntRx \equiv \left\{\tau_i: \arccos(\tau_i) \in \ElevRx  \right\} \subset [-1,1].$$ Define the width of the transmit and receive scattering intervals as
$
|\IntTx| = \int_{\IntTx}\!\!\!\!\! \, dt_j$ and $|\IntRx| = \int_{\IntRx}\!\!\!\!\! \, d\tau_i,
$ respectively.

\subsubsection{Hilbert space channel model}

Let $\TxSpace{j}$ be the Hilbert space of all square integrable transmit field distributions $X_j:\IntT{jj}\cup\IntT{ij}\rightarrow \mathbb{C}$  that transmitter $T_j$'s array of length $\LenTx$ can radiate in the direction of the available scattering clusters. 
%$\IntT{jj}\cup\IntT{ij}$ (both signal-of-interest and interference). %In the vernacular of \cite{TsePoon05DOF_EM}, $\TxSpace{j}$ is the space of field distributions array-limited to $\LenTx$ and wavevector-limited to $\IntT{jj}\cup\IntT{ij}$.
%To be precise, define $\TxSpace{j}$ to be the Hilbert space of all square-integrable functions  $X_j:\IntT{jj}\cup\IntT{ij}\rightarrow \mathbb{C}$, that can be expressed as
%$$
%X_j(t) = \int_{-\LenTx}^{\LenTx} \responseTx(t,p) x_j(p)\, dp,\ \quad t\in \IntT{jj}\cup\IntT{ij}
%$$
%for some $x_j(p),\ p\in [-\LenTx,\LenTx]$. 
The inner product between two member functions, $U_j,V_j\in\TxSpace{j}$, is the usual inner product
$
\langle U_j,V_j\rangle = \int_{\IntT{jj}\cup\IntT{ij}} U_j(t)V_j^*(t)\, dt.
$
Likewise let $\RxSpace{i}$ the Hilbert space of all received field distributions $Y_i:\IntR{ii}\cup\IntR{ij}\rightarrow \mathbb{C}$ incident on receiver $R_i$ and resolved by an array of length $\LenRx$.
%More precisely, $\RxSpace{i}$ is the Hilbert space of all square-integrable functions  $Y_i:\IntR{ii}\cup\IntR{ij} \rightarrow \mathbb{C}$, that can be expressed as  
%$$
%Y_i(\tau) = \int_{-\LenRx}^{\LenRx} \responseRx^*(q,\tau) y_i(q)\, dq,\ \quad \tau\in \IntR{ii}\cup\IntR{ij}
%$$
%for some $y_i(q),\ q\in [-\LenRx,\LenRx]$, with the usual inner product. 
From \cite{TsePoon05DOF_EM}, we know that the dimension of these transmit and receive signal spaces are, respectively, 
\begin{align}
\dim \TxSpace{j} &= 2\LenTx |\IntT{jj}\cup\IntT{ij}|, \\
\dim \RxSpace{i} &= 2\LenRx |\IntR{ii}\cup\IntR{ij}|.
\end{align}
%
%We can think of the scattering integrals in (\ref{eq:Zchan1}-\ref{eq:Zchan2}) as operators mapping from one Hilbert space to another.
 Define the channel scattering operator $\ScatOp{ij}:\TxSpace{j}\rightarrow\RxSpace{i}$ by
%\begin{equation}
%(\ScatOp{ij}X_j)(\tau) = \int_{\IntT{ij}} \Scat{ij}(\tau,t) X_j(t)\,dt, \qquad \tau \in \IntR{ij}
%\end{equation}
%\evanNote{Probably should be $\tau \in \IntR{ij}\cup \IntR{ii}$ to be more precise, only nonzero in $\IntR{ij}$.}
%\begin{equation}
%(\ScatOp{ij}X_j)(\tau) = \int_{\IntT{ij} \cup \IntT{jj}} \!\!\!\!\!\!\!\!\! \Scat{ij}(\tau,t) X_j(t)\,dt,\  \tau \in \IntR{ij}\cup \IntR{ii}.
%\end{equation}
\begin{equation}
(\ScatOp{ij}X_j)(\tau) = \int_{\IntT{ij} \cup \IntT{jj}} \!\!\!\!\!\!\!\!\! \Scat{ij}(\tau,t) X_j(t)\,dt,\  \tau \in \IntR{ij}\cup \IntR{ii}.
\end{equation}

With the above definitions, we write the channel input-output relationship 
\begin{align}
Y_1 &= \ScatOp{11}X_1 + \ScatOp{12}X_2 + Z_1,\\
Y_2 &= \ScatOp{22}X_2 + Z_2,
\end{align}
where $X_j \in \TxSpace{j}$ is the wavevector signal transmitted by $T_j$, $Y_i\in \RxSpace{i}$ is the wavevector signal received by $R_i$ and $Z_i \in \RxSpace{i}$ is additive noise. 
The impact of the scattering intervals is captured in the behavior of the scattering response integral kernel $\Scat{ij}(\tau,t)$, which we endow with the properties:
\begin{enumerate}

\item $\Scat{ij}(\tau,t) \neq 0$ only if $(\tau,t) \in \IntRx\times \IntTx$,
 
\item $\int||\Scat{ij}(\tau,t)||dt \neq 0\ \forall \ \tau \in \IntRx$,

\item $\int||\Scat{ij}(\tau,t)||d\tau \neq 0\ \forall\ t \in \IntTx$,

\item The point spectrum of $\Scat{ij}(\cdot,\cdot)$ is infinite.

\end{enumerate}
Let $R(\ScatOp{ij})\subset \RxSpace{i}$ denote the range of scattering operator $\ScatOp{ij}$, and let $R(\ScatOp{ij})^\perp \subset \RxSpace{i}$ denote the orthogonal complement of $R(\ScatOp{ij})$. Let $N(\ScatOp{ij})\subset \TxSpace{j}$ denote the nullspace of $\ScatOp{ij}$, and $N(\ScatOp{ij})^\perp$ its orthogonal space (i.e. the coimage of $\ScatOp{ij}$). The results of \cite{TsePoon05DOF_EM} can be combined with standard theorems of functional analysis to show the following properties:
\begin{align}
\dim R(\ScatOp{ij}) &= \dim N(\ScatOp{ij})^\perp \nonumber \\ &=  2\min\{\LenTx |\IntTx|, \LenRx |\IntRx| \}, \label{eq:prop1} \\
\dim N(\ScatOp{12}) &= 2\LenTwoTx |\IntT{22} \setminus \IntT{12}|  \nonumber \\
&\qquad+2(\LenTwoTx |\IntT{12}| - \LenOneRx |\IntR{12}|)^+,\label{eq:prop2}\\
\dim R(\ScatOp{11})^\perp &= 2\LenOneRx |\IntR{12} \setminus \IntR{11}| \nonumber \\
&\qquad+2(\LenOneRx |\IntR{11}| - \LenOneTx |\IntT{11}|)^+. \label{eq:prop3}
\end{align}

\section{Degrees-of-Freedom Analysis}

% !TEX root = ISIT_04_main.tex

\begin{thm}
\label{thm:mainResult}
Let $d_1$ and $d_2$ be the degrees of freedom of $\FlowOne$ and $\FlowTwo$ respectively. The degrees-of-freedom region, $\D$, of the three-node full-duplex channel is the convex hull of the degrees-of-freedom pairs, $(d_1,d_2)$, satisfying
\begin{align}
d_1 \leq &\ \dOneMax =  2\min ( \LenOneTx |\IntT{11}|, \LenOneRx |\IntR{11}| ), \label{eq:d1Bound}\\
d_2 \leq &\ \dTwoMax = 2 \min ( \LenTwoTx |\IntT{22}|, \LenTwoRx |\IntR{22}| ), \label{eq:d2Bound}\\
{d_1 + d_2} \leq  &\ \dSumMax =
2\LenTwoTx |\IntT{22} \setminus \IntT{12}| + 2\LenOneRx |\IntR{11} \setminus \IntR{12}|  \nonumber \\
& \qquad +2 \max (\LenTwoTx |\IntT{12}|,  \LenOneRx |\IntR{12}|) . \label{eq:dSumBound}
%\ \dSumMax \nonumber \\ &= \min \left\{ 
%% This first term is redundant given the above two terms 
%\vphantom{|\Cluster{11}{t}|} \right. 
%\min ( \LenOneTx |\IntT{11}|, \LenOneRx |\IntR{11}| ) \nonumber \\
% &\hspace{40pt} \qquad + \min ( \LenTwoTx |\IntT{22}|, \LenTwoRx |\IntR{22}|),  \nonumber \\
%% Term below is only active for an X-channel
%%&\hspace{40pt}   
%%\LenOneTx |\IntT{11} \setminus \IntT{21}| + \LenTwoRx |\IntR{22} \setminus \IntR{21}|  \nonumber \\
%%&\hspace{40pt} \qquad +\max (\LenOneTx |\IntT{21}|, \LenTwoRx |\IntR{21}|),  \nonumber \\
%%
%&\hspace{40pt}\LenTwoTx |\IntT{22} \setminus \IntT{12}| + \LenOneRx |\IntR{11} \setminus \IntR{12}|  \nonumber \\
%&\hspace{40pt} \qquad + \max (\LenTwoTx |\IntT{12}|, \left.  \LenOneRx |\IntR{12}|)   \right\}. \label{eq:dSumBound}
\end{align}
\end{thm}
The degrees-of-freedom region, $\D$, is depicted in Figure~\ref{fig:region}.
The achievability part of Theorem~\ref{thm:mainResult} is given in Section~\ref{sec:achieve} and a sketch of the converse is given in Section~\ref{sec:converse}. 
%\begin{IEEEproof}
%\ashuNote{Incorrect use of theorems. Theorem 2 proves achievability for Theorem 1, and Theorem 3 proves converse for Theorem 1. They should be just labeled achievability (direct part) and converse parts of the proof. You should say Achievability is in this subsection and Converse is that  subsection.}
%Theorem~\ref{thm:achieve} shows that all points $(d_1,d_2)$ within $\D$ are achievable, while Theorem~\ref{thm:converse} shows that $\D$ is also an outer bound on any achievable degrees-of-freedom region. %Thus $\D$ characterizes the set of every supportable degrees-of-freedom tuple. 
%\end{IEEEproof}

\begin{figure}[htbp]
\begin{center}
%\resizebox{0.5\textwidth}{!}{
	\begin{tikzpicture}
\begin{axis}[%
scale only axis,
width=2.2in,
height=1.2in,
xmin=0, xmax=6,
ymin=0, ymax=5,
xlabel={$d_1$},
ylabel={$d_2$},
ylabel near ticks,
xtick={1,2,3,4,5},
xticklabels={,{ \footnotesize $\dSumMax - \dOneMax$},,$\dOneMax$,},
ytick={1,2,3,4,5},
yticklabels={,{\footnotesize $\dSumMax - \dTwoMax$},,$\dTwoMax$,},
]
\addplot [color=blue, only marks, mark = *] coordinates{
	(2, 4)
	(4, 2)
};%
\addplot [color=black] coordinates{
	(4, 0)
	(4, 2)
	(2, 4)
	(0, 4)
};
\addplot [color=black, dashed] coordinates{
	(2, 0)
	(2, 4)
};%
\addplot [color=black, dashed] coordinates{
	(0, 2)
	(4, 2)
};%
\node[pin=60:{\small $d_1+d_2 = \dSumMax$}] at (axis description cs:0.50,0.55) {};
\node[] at (axis description cs:0.44,0.87) {\small$(d_1'', d_2'')$};
\node[] at (axis description cs:0.78,0.4) {\small$(d_1', d_2')$};
\end{axis}
\end{tikzpicture}
%}
\caption{degrees-of-freedom region, $\D$}
\label{fig:region}
\end{center}
\end{figure}

	\subsection{Achievability}
	
	% !TEX root = ISIT_04_main.tex

%Tthis is the direct part of Theorem 1}
\label{sec:achieve}
We establish achievability of $\D$ by way of two lemmas. The first lemma shows the achievability of two specific degree-of-freedom pairs, and the second lemma remarks that these pairs are the corner points of $\D$. 

\begin{lem}
%\label{thm:ZChanOne}
\label{thm:achievePoints}
The degree-of-freedom pairs $(d_1',d_2')$ and $(d_1'',d_2'')$ are achievable, where 
\begin{align}
d_1' =&\min\left\{2\LenOneTx |\IntT{11}|, 2\LenOneRx |\IntR{11}|\right\}, \label{eq:point1} \\
d_2' =& \min\left\{d_{T_2},2\LenTwoRx |\IntR{22}|\right\} 1(\LenOneTx |\IntT{11}| \geq \LenOneRx |\IntR{11}|) \nonumber \\ 
 &+ \min\left\{\delta_{T_2},2\LenTwoRx |\IntR{22}|\right\} 1(\LenOneTx |\IntT{11}| < \LenOneRx |\IntR{11}|), \label{eq:point2}\\
d_1'' =& \min\left\{2\LenOneTx |\IntT{11}|, d_{R_1}\right\} 1(\LenTwoRx |\IntR{22}| \geq \LenTwoTx |\IntT{22}|) \nonumber \\ 
 &+ \min\left\{2\LenOneTx |\IntT{11}|, \delta_{R_1} \right\} 1(\LenTwoRx |\IntR{22}| < \LenTwoTx |\IntT{22}|),\\
d_2'' =&\min\left\{2\LenTwoTx |\IntT{22}|, 2\LenTwoRx |\IntR{22}|\right\}, 
\end{align}
with $d_{T_2}$,$\delta_{T_2}$, $d_{R_1}$, and $\delta_{R_1}$ given in (\ref{eq:dT2}-\ref{eq:deltaR1}) at the top of the following page. 
%
%The degree-of-freedom pairs $(d_1',d_2')$ is achievable, where 
%\begin{align}
%d_1' =&\min\left\{2\LenOneTx |\IntT{11}|, 2\LenOneRx |\IntR{11}|\right\}, \\
%d_2' =& \min\left\{d_{T_2},2\LenTwoRx |\IntR{22}|\right\} 1(\LenOneTx |\IntT{11}| \geq \LenOneRx |\IntR{11}|) \nonumber \\ 
% &+ \min\left\{\delta_{T_2},2\LenTwoRx |\IntR{22}|\right\} 1(\LenOneTx |\IntT{11}| < \LenOneRx |\IntR{11}|),
%\end{align}
%where $d_{T_2}$ and $\delta_{T_2}$ are given in (\ref{eq:dT2}-\ref{eq:deltaT2}) at the top of the following page. Likewise, the degree-of-freedom pair $(d_1'',d_2'')$ is achievable, where
%\begin{align}
%d_1'' =& \min\left\{2\LenOneTx |\IntT{11}|, d_{R_1}\right\} 1(\LenTwoRx |\IntR{22}| \geq \LenTwoTx |\IntT{22}|) \nonumber \\ 
% &+ \min\left\{2\LenOneTx |\IntT{11}|, \delta_{R_1} \right\} 1(\LenTwoRx |\IntR{22}| < \LenTwoTx |\IntT{22}|),\\
%d_2'' =&\min\left\{2\LenTwoTx |\IntT{22}|, 2\LenTwoRx |\IntR{22}|\right\}, 
%\end{align}
%where $d_{R_1}$ and $\delta_{R_1}$ are given in (\ref{eq:dR1}-\ref{eq:deltaR1}) at the bottom of the following page. 
%
\begin{figure*}[!t]
% ensure that we have normalsize text
\normalsize
% Store the current equation number.
\setcounter{MYtempeqncnt}{\value{equation}}
% Set the equation number to one less than the one
% desired for the first equation here.
% The value here will have to changed if equations
% are added or removed prior to the place these
% equations are referenced in the main text.
\setcounter{equation}{16}
\hrulefill
\begin{align}
d_{T_2} =& 2 \LenTwoTx |\IntT{22} \setminus \IntT{12}|
+ 2\min \left\{ \vphantom{\left(|\IntT{12}|\right)^+}  \LenTwoTx |\IntT{22} \cap \IntT{12}|,  
(\LenTwoTx |\IntT{12}| - \LenOneRx |\IntR{12}| )^+  +   \LenOneRx |\IntR{12} \setminus \IntR{11}| \right\}^+ \label{eq:dT2}\\
%
%d_{R_2} =& \LenTwoRx |\IntR{22} \setminus \IntR{12}| + \min \left\{ {\scriptstyle \vphantom{\left(|\IntT{12}|\right)^+} \LenTwoRx |\IntR{22} \cap \IntR{12}|,
%\LenTwoRx |\IntR{21}| - \left[\LenOneRx|\IntR{11}| - \left(\LenOneTx \IntT{11} \setminus \IntT{21} + (\LenOneTx |\IntT{21}| - \LenTwoRx |\IntR{21}|)^+ \right) \right]^+ } \right\} \label{eq:d2r}\\
%
\delta_{T_2} =& 2\LenTwoTx |\IntT{22} \setminus \IntT{12}| + 2\min \left\{ {\scriptstyle\vphantom{\left(|\IntR{12}|\right)^+} \LenTwoTx |\IntT{22} \cap \IntT{12}|,\ 
\LenTwoTx |\IntT{12}| - \left[\LenOneTx|\IntT{11}| - \left(\LenOneRx |\IntR{11} \setminus \IntR{12}| + (\LenTwoRx |\IntR{12}| - \LenOneTx |\IntT{12}|)^+ \right) \right]^+ } \right\} \label{eq:deltaT2}
%
%\\ \delta_{R_2} =& \LenTwoRx |\IntR{22} \setminus \IntR{21}| + \min\left\{\LenTwoRx|\IntR{22} \cap \IntR{21}|,
%(\LenTwoRx|\IntR{21}| - \LenOneTx|\IntT{21}|)^+ + \LenOneTx|\IntT{21} \setminus \IntT{11}| \right\}^+\label{eq:delta2r}
\end{align}
\begin{align}
d_{R_1} =& 2\LenOneRx |\IntR{11} \setminus \IntR{12}| + 2\min\left\{\LenOneRx|\IntR{11} \cap \IntR{12}|,
(\LenOneRx|\IntR{12}| - \LenTwoTx|\IntT{12}|)^+ + \LenTwoTx|\IntT{12} \setminus \IntT{22}|\right\}^+\label{eq:dR1}
\\
\delta_{R_1} =& 2 \LenOneRx |\IntR{11} \setminus \IntR{12}| + 2\min \left\{ {\scriptstyle \vphantom{\left(|\IntR{21}|\right)^+} \LenOneRx |\ClusterR{11} \cap \IntR{12}|,\ 
%\right. \nonumber \\  & \left. \hspace{200pt} 
\LenOneRx |\IntR{12}| - \left[\LenTwoRx|\IntR{22}| - \left(\LenTwoTx |\IntT{22} \setminus \IntT{12}| + (\LenTwoTx |\IntT{12}| - \LenOneRx |\IntR{12}|)^+ \right) \right]^+} \right\} \label{eq:deltaR1}
\end{align}
% Restore the current equation number.
\setcounter{equation}{\value{MYtempeqncnt}}
% IEEE uses as a separator
\hrulefill
% The spacer can be tweaked to stop underfull vboxes.
\vspace*{4pt}
\end{figure*}
\end{lem}
\setcounter{equation}{20}
	\begin{IEEEproof}[Sketch of Proof]
%The full proof cannot be included within the page constraint, but
The proof is inspired by the zero-forcing scheme of \cite{Jafar07MIMOInterferenceDoF} for the MIMO interference channel, except that processing is performed in continuous Hilbert spaces rather than discrete vector spaces, and the fact that scattering intervals are not perfectly overlapped requires some extra treatment. 
The full proof is omitted for brevity, but sketch the  achievability of $(d_1',d_2')$ when 
$\LenOneTx |\IntT{11}| \geq \LenOneRx |\IntR{11}|$, 
$d_{T_2}\leq 2\LenTwoRx |\IntR{22}|$,  
$\LenTwoTx |\IntT{22} \cap \IntT{12}| \geq 2(\LenTwoTx |\IntT{12}| - \LenOneRx |\IntR{12}| )^+ + 2\LenOneRx |\IntR{12} \setminus \IntR{11}|$, 
and 
$\LenTwoTx |\IntT{12}| \geq \LenOneRx |\IntR{12}|$.
In this case (\ref{eq:point1})~and~(\ref{eq:point2}) simplify to
\begin{align}
d_1' &=2\LenOneRx |\IntR{11}|, \label{eq:d1PrimeCase1} \\
d_2' & \label{eq:d2PrimeCase1}
=2( \LenTwoTx |\IntT{22} \setminus \IntT{12}| +  (\LenTwoTx |\IntT{12}| - \LenOneRx |\IntR{12}|) \nonumber   \\
&\quad +  \LenOneRx |\IntR{12} \setminus \IntR{11}|).
\end{align}

We give $\FlowOne$ its maximum point-to-point degrees of freedom, which is shown in \cite{TsePoon05DOF_EM} to be $\min\{2\LenOneTx |\IntT{11} ,2\LenOneRx |\IntR{11}|\} =  2\LenOneRx |\IntR{11}| = d_1'$. 
The wavevector received at $R_1$ from $T_1$, $\ScatOp{11}X_1$, necessarily lies in $R(\ScatOp{11})$. If $T_2$ can construct its transmitted wavevector signal, $X_2$, such that 
%\begin{equation}
$
\ScatOp{12}X_2\in R(\ScatOp{11})^\perp
$
%\end{equation} 
then we will have $\ScatOp{11}X_1\perp\ScatOp{12}X_2$
and thus $T_2$ will not impede $R_1$'s recovery of the $d_1'$ symbols from $T_1$. 
Let
%\begin{equation}
$
\TxTwoPreSpace \equiv \preim{R(\ScatOp{11})^\perp}{\ScatOp{12}} \subseteq \TxSpace{2} 
$
%\end{equation}
denote the preimage of $R(\ScatOp{11})^\perp$ under $\ScatOp{12}$.
Then constructing $X_2$ such that $X_2\in\TxTwoPreSpace$ ensures $\ScatOp{11}X_1\perp\ScatOp{12}X_2$.
Since we are considering the case where $\LenTwoTx |\IntT{12}| \geq \LenOneRx |\IntR{12}|$, $R(\ScatOp{11})^\perp\subseteq R(\ScatOp{12})$
and thus 
\begin{align}
\dim \TxTwoPreSpace &= \dim N(\ScatOp{12}) + \dim R(\ScatOp{11})^\perp\\
					&=2( \LenTwoTx |\IntT{22} \setminus \IntT{12}| +  (\LenTwoTx |\IntT{12}| - \LenOneRx |\IntR{12}|) \nonumber   \\
&\quad +  \LenOneRx |\IntR{12} \setminus \IntR{11}|) \label{eq:props}\\
					&=d_2',
\end{align}
where in (\ref{eq:props}) we have leveraged properties (\ref{eq:prop2}) and (\ref{eq:prop3}) from Section~\ref{sec:systemModel}.
Therfore $T_2$ can transmit the required $d_2'$ symbols along each basis function of any orthonormal basis of $\TxTwoPreSpace$, thus avoiding interfering $R_1$.
And since in the case we are considering $d_2'$ is no larger than $\min\left\{2\LenTwoTx |\IntT{22}|, 2\LenTwoRx |\IntR{22}|\right\}$, which is the number of degrees of freedom  $\FlowTwo$ can support, $R_2$ can recover the $d_2'$ of the symbols transmitted from $T_2$, as desired. 
\end{IEEEproof}

	\begin{lem}
\label{lem:corners}
%The degree-of-freedom pairs $(d_1',d_2')$ and $(d_1'',d_2'')$, are equal to the corner points of the degree-of-freedom region, $\D$, defined in Theorem~\ref{thm:mainResult}. More precisely, 
%\begin{align}
%(d_1',d_2') &= (\dOneMax, \dSum - \dOneMax) \label{eq:corner1} \\
%(d_1'',d_2'') &= (\dSum - \dTwoMax,  \dTwoMax) \label{eq:corner2}
%%\left( \min ( \SizeOneTx |\Cluster{11}{t}|, \SizeOneRx |\Cluster{11}{r}|),
%%D_{\sf sum}^{\sf max} - \min ( \SizeOneTx |\Cluster{11}{t}|, \SizeOneRx |\Cluster{11}{r}|)  \right)
%\end{align}
The degree-of-freedom pairs $(d_1',d_2')$ and $(d_1'',d_2'')$, are the corner points of $\D$.
\end{lem}

	\begin{IEEEproof}[Sketch of Proof]
One can check that
\begin{align}
(d_1',d_2') &= (\dOneMax, \dSum - \dOneMax) \label{eq:corner1} \\
(d_1'',d_2'') &= (\dSum - \dTwoMax,  \dTwoMax) \label{eq:corner2}.
\end{align}
by exhausting computing the left and right and sides of (\ref{eq:corner1})~and~(\ref{eq:corner2}) in all cases and observing equality. We omit the computations for brevity. 
%Equations (\ref{eq:corner1})~and~(\ref{eq:corner2}) can be verified by computing the left- and right-hand sides for all combinations of the conditions 
%$\LenOneTx|\IntT{11}|  \lesseqgtr  \LenOneRx|\IntR{11}|$,
%$\LenTwoTx|\IntT{12}|  \lesseqgtr  \LenOneRx|\IntR{12}|$,
%and $\LenTwoTx|\IntT{22}|  \lesseqgtr  \LenTwoRx|\IntR{22}|$
%%\begin{align}
%%\LenOneTx|\IntT{11}|  &\lesseqgtr  \LenOneRx|\IntR{11}|,\\
%%%\LenOneTx|\IntT{21}|  &\lesseqgtr  \LenTwoRx|\IntR{21}|\\
%%\LenTwoTx|\IntT{12}|  &\lesseqgtr  \LenOneRx|\IntR{12}|,\\
%%\LenTwoTx|\IntT{22}|  &\lesseqgtr  \LenTwoRx|\IntR{22}|
%%\end{align}
%and observing equality in each of the $2^3 = 8$ cases. 
\end{IEEEproof}
	
	%\begin{thm}
%\label{thm:achieve}
%Every degree-of-freedom pair $(d_1,d_2)\in \D$ is achievable.
%\end{thm}
%\ashuNote{this theorem statement is not needed}
%
%\begin{proof}
%Lemmas~\ref{thm:achievePoints} and \ref{lem:corners} show that the corner points of $\D$, $(d_1',d_2')$ and $(d_1'',d_2'')$ are achievable. And thus all other points within $\D$ are achievable via time sharing between the schemes that achieve the corner points. 
%%
%%All other points along the line between the points $(d_1',d_2')$ and $(d_1'',d_2'')$ are thus achievable via time sharing between the schemes that achieve each conrner. All other points in the region require less degrees of freedom than the points along the line from $(d_1',d_2')$ and $(d_1'',d_2'')$ and are thus trivially achievable. 
%\end{proof}
Lemmas~\ref{thm:achievePoints} and \ref{lem:corners} show that the corner points of $\D$, $(d_1',d_2')$ and $(d_1'',d_2'')$ are achievable. And thus all other points within $\D$ are achievable via time sharing between the schemes that achieve the corner points.

	\subsection{Converse}
	\label{sec:converse}
%\begin{thm}
%\label{thm:converse}
%If the degree-of-freedom pair $(d_1,d_2)$ is achievable, then $(d_1,d_2) \in \D$.
%\end{thm}
%\begin{IEEEproof}[Sketch of Converse]
%\end{IEEEproof}

%\ashuNote{remove theorem statement and just state the sketch of converse}
The full converse is omitted, for brevity, but here we give a sketch of the procedure for showing the converse part of Theorem~\ref{thm:mainResult}.
We would like to show that if the degree-of-freedom pair $(d_1,d_2)$ is achievable, then $(d_1,d_2) \in \D$.
%\evanNote{Finish this} 
It is easy to see that if $(d_1,d_2)$ is achievable, then constraints (\ref{eq:d1Bound}) and (\ref{eq:d2Bound}) must be satisfied as these are the point-to-point bounds given in \cite{TsePoon05DOF_EM}. 
It remains to show that the sum degree-of-freedom constrain (\ref{eq:dSumBound}) must hold for every achievable $(d_1,d_2)$. 
Our process for showing (\ref{eq:dSumBound}) is twofold. %First, a genie expands each of the scattering intervals until the interference and signal-of-interest intervals are overlapping and identical. The genie also lengthens each of the arrays such that the any added interference due to the overlapping scattering intervals is compensated so that the net manipulation of the genie only enlarge $\D$. 

First, a genie expands the scattering intervals $\IntT{22}$ and $\IntT{12}$, to $\IntTPrime{22} = \IntTPrime{12} = \IntT{22} \cup \IntT{12}$,  and also expands $\IntR{11}$ and $\IntR{12}$ to  $\IntRPrime{11} = \IntRPrime{12} = \IntR{11} \cup \IntR{12}$. The genie also lengthens the $T_2$ array to $\LenTwoTxPrime = \LenTwoTx +  \LenOneRx \frac{|\IntR{11} \setminus \IntR{12}|} {|\IntT{22} \cup \IntT{12}|}$ and the $R_1$ array to length $\LenOneRxPrime = \LenOneRx +  \LenTwoTx \frac{|\IntT{22} \setminus \IntT{12}|} {|\IntR{11} \cup \IntR{12}|},$
which one can show ensures that any added interference due to the expansion of $\IntT{12}$ and $\IntR{12}$ is compensated by the larger arrays sizes so that the net manipulation of the genie can only enlarge $\D$. 

%each of the scattering intervals until the interference and signal-of-interest intervals are overlapping and identical. The genie also lengthens each of the arrays such that the any added interference due to the overlapping scattering intervals is compensated so that the net manipulation of the genie only enlarge $\D$. 

One can check that after above genie manipulation is performed, the maximum of the $T_2$ and $R_1$ signaling dimensions are equal to $\dSumMax$ in constraint (\ref{eq:dSumBound}), and
since the scattering intervals are overlapped, the channel model becomes the Hilbert space equivalent of the well-studied MIMO $Z$-channel. The Hilbert space analog to the bounding techniques employed in \cite{{Ke12ZDoF,Jafar07MIMOInterferenceDoF}} that show the sum degrees of freedom of the MIMO $z$-channel is bound by $\max(M_2,N_1)$ can be leveraged to show (\ref{eq:dSumBound}) as desired. 

%First, let a genie increase $\LenOneTx$ and $\LenTwoRx$, the aperture size of the $T_1$ and $R_2$ arrays, towards infinity, such that the sum degrees of freedom becomes limited only by the signaling dimensions available at $T_2$ and $R_1$. Clearly, this genie will only increase the sum degrees of freedom.  

%Combining Theorems~\ref{thm:achieve}~and~\ref{thm:converse} establishes Theorem 1. 

%\section{Relation to MIMO $Z$-Channel}

\section{Impact on Full-duplex Design}
Let $\DHD$ be the region of degree-of-freedom pairs achievable via half-duplex mode, i.e. by time-division-duplex between transmission at $T_1$ and $T_2$, so that there is no self-interference in this case.  It is easy to see that the half-duplex achievable region is characterized by
\begin{align}
d_1 &\leq \alpha \min\left\{2\LenOneTx |\IntT{11}|, 2\LenOneRx |\IntR{11}|\right\},\\
d_2 & \leq (1-\alpha)\min\left\{2\LenTwoTx |\IntT{22}|, 2\LenTwoRx |\IntR{22}|\right\},
\end{align}
where $\alpha \in [0,1]$ is the time sharing parameter.  
Obviously $\DHD\subseteq \D$, but we are interested in contrasting the scenarios for which $\DHD\subset\D$, and full-duplex spatial isolation strictly outperforms half-duplex time division, and the scenarios for which $\DHD=\D$ and half-duplex can achieve the same performance as full-duplex. We will consider two particularly interesting cases: the fully spread environment, and the symmetric spread environment. 

\subsection{Fully Spread}
Consider case where the environment is fully spread,
$$
{|\IntT{11}| =  |\IntR{11}| = |\IntT{22}| = |\IntR{22}| = |\IntT{12}|= |\IntR{12}| = 2.}
$$
For simplicity also assume that the base station transmit and receive arrays are of length $\LenOneRx = \LenTwoTx = \LenBS$, and user arrays are of length $\LenOneTx = \LenTwoRx = \LenU$. In this case the full-duplex degree-of-freedom region, $\D$, simplifies to 
\begin{align}
d_i \leq 4\min\{\LenBS,\LenU\},i=1,2;\quad \label{eq:fullFD} 
d_1 + d_2 \leq 4\LenBS %\label{eq:fullSum}
\end{align}
while the half-duplex achievable region, $\DHD$ simplifies to
\begin{align}
d_1 + d_2 \leq 4\min\{\LenBS,\LenU\}. \label{eq:fullHD} 
\end{align}
\begin{rem}
In the fully-scattered case, $\DHD \subset \D$ if $\LenBS > \LenU$, else $\DHD = \D$. 
\end{rem}
%The intuition is that full-duplex will be better than half-duplex when the base station arrays are larger than the user arrays such that are left for null-steering even after maximizing spatial multiplexing to users. \ashuNote{sentence needs to be rewritten - does not make sense}

%
%
%\evanNote{Coming soon if space/time permit... This is the case that can be shown analogous to the MIMO $z$-channel. And here FD will be better than HD when the base station arrays are larger than the user arrays, since resources are left for null-steering even after maximizing the spatial multiplexing to users}

\subsection{Symmetric Spread}
We will consider a special case that illustrates the impact of the overlap of the scattering intervals on full-duplex performance. 
Assume all the arrays in the network, the two arrays on the base station as well as the array on each of the user devices, are of the same length $L$, that is 
$\LenOneTx =  \LenOneRx = \LenTwoTx = \LenTwoRx = L.$
Assume also the size of the scattering interval to/from the intended receiver/transmitter is the same for all arrays
$|\IntT{11}| =  |\IntR{11}| = |\IntT{22}| = |\IntR{22}| = |\IntFwd|.$  Finally assume that
$ |\IntT{12}| = |\IntR{12}| = |\IntBack|,$
and that the amount of overlap with the intended-signal scattering interval is the same so that
%\begin{align}
%|\IntT{22}\setminus\IntT{12}| = |\IntR{11}\setminus\IntR{12}| &=  |\IntFwd \setminus \IntBack|.
%\end{align}
%\begin{align}
%|\IntT{22}\cap\IntT{12}| = |\IntR{11}\cap\IntR{12}| &=  |\IntFwd \cap \IntBack|\\
%& = |\IntFwd| - |\IntFwd \setminus \IntBack|.
%\end{align}
$
|\IntT{22}\cap\IntT{12}| = |\IntR{11}\cap\IntR{12}| =  |\IntFwd \cap \IntBack| = |\IntFwd| - |\IntFwd \setminus \IntBack|.
$

%We call $\IntFwd$ the \emph{forward} interval, since from the base stations point of view, it is the set of directions in which signal scatter to/from  the intended 
We call $\IntBack$ the \emph{backscatter interval} since it is the angle subtended at the base station by the back-scattering clusters, while we call  $\IntFwd$ the \emph{forward interval}, since it is the angle subtended by the clusters that scatter towards the intended transmitter/receiver. 
In this symmetric case, the full-duplex degree-of-freedom region, $\D$ simplifies to 
\begin{align}
d_i &\leq 2L|\IntFwd|,\ i=1,2 \label{eq:symInd}\\
d_1 + d_2 &\leq 2L(2 |\IntFwd\setminus\IntBack| + |\IntBack|) \label{eq:symSum}
\end{align}
while the half-duplex achievable region, $\DHD$ is 
\begin{align}
d_1 + d_2 &\leq 2L|\IntFwd|.
\end{align}

\begin{rem}
%Comparing $\D$ and $\DHD$ above we see that $L\neq0$, 
Comparing $\D$ and $\DHD$ above we see that in the case of symmetric scattering, $\DHD = \D$ if and only if $\IntFwd=\IntBack$,\footnote{We are neglecting the trivial case of $L=0$.} else $\DHD \subset \D$.
\end{rem}
%
%It is easy to see that $\DHD = \D$ if and only if $\IntFwd=\IntBack$,\footnote{neglecting the trivial case of $L=0$} else $\DHD \subset \D$. 
Thus the full-duplex spatial isolation region is strictly larger than the half-duplex time-division region unless the forward interval and the backscattering interval are exactly aligned. The intuition is that when $\IntFwd=\IntBack$ the scattering interval must be shared, just as time must be, thus trading spatial resources is equivalent to trading time-slots. However, if $\IntFwd\neq\IntBack$, there is a portion of space exclusive to each user, and can be leveraged to improve upon time division. 

%Inspection of $\D$ above leads to the following remark 
\begin{rem}
In the case of symmetric scattering, the full-duplex degree-of-freedom region is rectangular if and only if
\begin{equation}
|\IntBack\setminus\IntFwd| \geq |\IntFwd\cap\IntBack| \label{eq:sqCond}.
\end{equation}
\end{rem} 
The above remark can be verified by  comparing (\ref{eq:symInd}) and (\ref{eq:symSum}) observing that the sum-rate bound, (\ref{eq:symSum}),  is only active when 
\begin{equation}
2|\IntFwd\setminus\IntBack| + |\IntBack|  \geq 2|\IntFwd|. \label{eq:cond1}
\end{equation}
A few lines of set-algebraic manipulation of condition (\ref{eq:cond1}) shows that it is equivalent to (\ref{eq:sqCond}). 
%\begin{IEEEproof}
%By comparing (\ref{eq:symInd}) and (\ref{eq:symSum}) we see that the sum-rate bound is only active when $2|\IntFwd\setminus\IntBack| + |\IntBack|  \geq 2|\IntFwd|$ which the following algebraic steps show is equivalent to the condition given in the remark. 
%\begin{align}
%2|\IntFwd\setminus\IntBack| + |\IntBack|  \geq 2|\IntFwd| \\
%2|\IntFwd| -2|\IntFwd\cap\IntBack| + |\IntBack|  \geq 2|\IntFwd| \\
%2|\IntFwd\cap\IntBack| \leq   |\IntBack| \\
%|\IntBack|-|\IntBack\setminus\IntFwd|+|\IntFwd\cap\IntBack| \leq  |\IntBack| \\ 
%|\IntBack\setminus\IntFwd| \geq |\IntFwd\cap\IntBack|
%\end{align}
%\end{IEEEproof}
One intuition behind this remark is that when $|\IntBack\setminus\IntFwd| \geq |\IntFwd\cap\IntBack|$, then the interval $|\IntFwd\cap\IntBack|$ can be used as interference free side-channel on which it can communicate the interference it is generating over $|\IntFwd\cap\IntBack|$, so that the interference can be cancelled. %\evanNote{I need another, more spatial intuition, as this raises the question ``why is this not already canceled since it's at the same node?''}

\begin{figure}[htbp]
\begin{center}
\begin{tikzpicture}
\begin{axis}[%
scale only axis,
width=2in,
height=1.4in,
xmin=0, xmax=2.5,
ymin=0, ymax=2.5,
xlabel={$d_1$},
ylabel={$d_2$},
ylabel near ticks,
xtick={.5,1,1.5,2, 2.5, 3},
xticklabels={$\frac{L}{2}$,$L$,$\frac{3L}{2}$,$2L$},
ytick={.5,1,1.5,2},
yticklabels={$\frac{L}{2}$,$L$,$\frac{3L}{2}$,$2L$},
%axis on top,
legend entries={
{\footnotesize$\DHD$},
{\footnotesize$\D$: $|\IntFwd\cap\IntBack| = 1$},
{\footnotesize$\D$: $|\IntFwd\cap\IntBack| = 0.75$},
{\footnotesize$\D$: $|\IntFwd\cap\IntBack| \leq 0.5$}
},
legend style={ at={(0.5,1.1)},anchor=center,nodes=right}]
%legend style={fill = KeynoteBG, at={(0.05,0.02)},anchor=south west,nodes=right}]
%
%\addplot [
%color=KeynoteGray,dashed,line width=2.0pt] 
%table [x = gamma_dB, y = R_HD_PeakPwr]{Rate_vs_SI.dat};
%
\addplot [color=black, line width=2.0pt] coordinates{
	(2, 0)
	(0, 2)
};%
\addplot [color=green, line width=2.0pt, dotted] coordinates{
	(2, 0)
	(0, 2)
};%
\addplot [color=red, line width=2.0pt] coordinates{
	(2, 0)
	(2, 1)
	(1, 2)
	(0, 2)
};%
\addplot [color=blue, line width=2.0pt, dashed] coordinates{
	(2, 0)
	(2, 2)
	(0, 2)
};%
\end{axis}
\end{tikzpicture}
\caption{Symmetric-spread degree-of-freedom regions for different amounts of scattering overlap}
\label{fig:symRegions}
\end{center}
\end{figure}
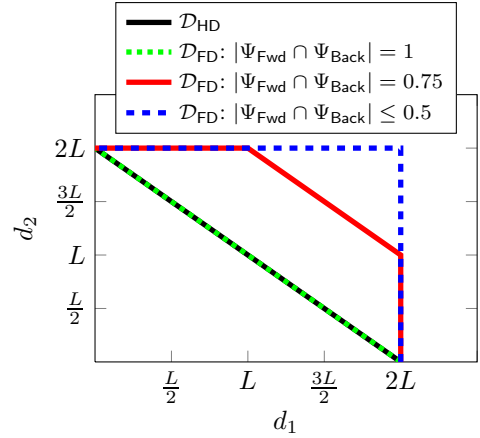

Consider the case where $|\IntFwd| = 1$ and $|\IntBack| = 1$, thus the overlap between the two, $|\IntFwd\cap\IntBack|$ can vary from zero to one. Figure~\ref{fig:symRegions} plots the half-duplex region, $\DHD$, and the full-duplex region, $\D$, for several different values  of overlap, $|\IntFwd\cap\IntBack|$. We see that when $\IntFwd=\IntBack$ so that $|\IntFwd\cap\IntBack|=1$, both $\DHD$ and $\D$ are the same triangular region. When $|\IntFwd\cap\IntBack|=0.75$, we get a rectangular region. Once $|\IntFwd\cap\IntBack|\leq 0.5$, $|\IntBack\setminus\IntFwd|$ becomes greater than 0.5, such that condition of (\ref{eq:sqCond}) is satisfied and the degree-of-freedom region becomes rectangular.

%
%since it is the set of directions in which the base station transmitter, $T_2$, can transmit will that scatter to the intended downlink user, $R_2$. 

%\section{Conclusion}

%\iftoggle{ZChanModel}{
%\section{Results for Unipolarized Linear Arrays}
%}{
%\section{Results for Z-Channel with Unipolarized Linear Arrays}
%Consider the case where there is no coupling from $T_1$ to $R_2$, i.e. the Z-channel. When we have unipolarized linear arrays, a Z-channel occurs whenever $\IntT{21} = \IntR{21} = \emptyset$. In this case we can write the wave-vector domain channel model as  
%\begin{align}
%Y_1(\tau) &=  \int_{\IntT{11}} \!\!\!\!\!\!\!\! H_{11}(\tau,t) X_1(t)\, dt
%			+ \int_{\IntT{12}} \!\!\!\!\!\!\!\! H_{12}(\tau,t) X_2(t)\, dt, 			\label{eq:Zchan1}
%			\\
%Y_2(\tau) &=  \int_{\IntT{22}} \!\!\!\!\!\!\!\! H_{22}(\tau,t) X_2(t)\, dt\label{eq:Zchan2}.
%\end{align}
%}
%\subsection{Achievability}
%\input{ZChanUnipolAchieve}

\ifCLASSOPTIONcaptionsoff
  \newpage
\fi

\bibliographystyle{IEEEtran}
\bibliography{IEEEabrv,/Users/evaneverett/Dropbox/Wireless_Literature/Bibliography/Research}

% Generated by IEEEtran.bst, version: 1.13 (2008/09/30)
\begin{thebibliography}{10}
\providecommand{\url}[1]{#1}
\csname url@samestyle\endcsname
\providecommand{\newblock}{\relax}
\providecommand{\bibinfo}[2]{#2}
\providecommand{\BIBentrySTDinterwordspacing}{\spaceskip=0pt\relax}
\providecommand{\BIBentryALTinterwordstretchfactor}{4}
\providecommand{\BIBentryALTinterwordspacing}{\spaceskip=\fontdimen2\font plus
\BIBentryALTinterwordstretchfactor\fontdimen3\font minus
  \fontdimen4\font\relax}
\providecommand{\BIBforeignlanguage}[2]{{%
\expandafter\ifx\csname l@#1\endcsname\relax
\typeout{** WARNING: IEEEtran.bst: No hyphenation pattern has been}%
\typeout{** loaded for the language `#1'. Using the pattern for}%
\typeout{** the default language instead.}%
\else
\language=\csname l@#1\endcsname
\fi
#2}}
\providecommand{\BIBdecl}{\relax}
\BIBdecl

\bibitem{Duarte11FullDuplex}
M.~Duarte, C.~Dick, and A.~Sabharwal, ``Experiment-driven characterization of
  full-duplex wireless systems,'' \emph{Wireless Communications, IEEE
  Transactions on}, vol.~11, no.~12, pp. 4296--4307, 2012.

\bibitem{SahaiPhaseNoiseDraft}
A.~Sahai, G.~Patel, C.~Dick, and A.~Sabharwal, ``On the impact of phase noise
  on active cancelation in wireless full-duplex,'' \emph{Vehicular Technology,
  IEEE Transactions on}, vol.~62, no.~9, pp. 4494--4510, 2013.

\bibitem{Bliss07SimultTX_RX}
D.~W. Bliss, P.~A. Parker, and A.~R. Margetts, ``Simultaneous transmission and
  reception for improved wireless network performance,'' in \emph{Proceedings
  of the 2007 IEEE/SP 14th Workshop on Statistical Signal Processing}.\hskip
  1em plus 0.5em minus 0.4em\relax Washington, DC, USA: IEEE Computer Society,
  2007, pp. 478--482.

\bibitem{Day12FDMIMO}
B.~Day, A.~Margetts, D.~Bliss, and P.~Schniter, ``Full-duplex bidirectional
  {MIMO}: Achievable rates under limited dynamic range,'' \emph{Signal
  Processing, IEEE Transactions on}, vol.~60, no.~7, pp. 3702 --3713, july
  2012.

\bibitem{Everett11Asilomar}
E.~Everett, M.~Duarte, C.~Dick, and A.~Sabharwal, ``Empowering full-duplex
  wireless communication by exploiting directional diversity,'' in
  \emph{Asilomar Conference on Signals, Systems and Computers}, October 2011.

\bibitem{Everett2013PassiveSuppressionFD}
\BIBentryALTinterwordspacing
E.~Everett, A.~Sahai, and A.~Sabharwal, ``Passive self-interference suppression
  for full-duplex infrastructure nodes,'' \emph{Wireless Communications, IEEE
  Transactions on}, to be published. [Online]. Available:
  \url{http://arxiv.org/abs/1302.2185}
\BIBentrySTDinterwordspacing

\bibitem{TsePoon05DOF_EM}
A.~Poon, R.~Brodersen, and D.~Tse, ``Degrees of freedom in multiple-antenna
  channels: a signal space approach,'' \emph{Information Theory, IEEE
  Transactions on}, vol.~51, no.~2, pp. 523 -- 536, feb. 2005.

\bibitem{TsePoon06EmagInfoTheory}
A.~Poon, D.~Tse, and R.~Brodersen, ``Impact of scattering on the capacity,
  diversity, and propagation range of multiple-antenna channels,''
  \emph{Information Theory, IEEE Transactions on}, vol.~52, no.~3, pp. 1087
  --1100, march 2006.

\bibitem{TsePoon11DOFPolarization}
A.~Poon and D.~Tse, ``Degree-of-freedom gain from using polarimetric antenna
  elements,'' \emph{Information Theory, IEEE Transactions on}, vol.~57, no.~9,
  pp. 5695 --5709, sept. 2011.

\bibitem{Jafar07MIMOInterferenceDoF}
S.~Jafar and M.~Fakhereddin, ``Degrees of freedom for the mimo interference
  channel,'' \emph{Information Theory, IEEE Transactions on}, vol.~53, no.~7,
  pp. 2637--2642, 2007.

\bibitem{Ke12ZDoF}
L.~Ke and Z.~Wang, ``Degrees of freedom regions of two-user mimo z and full
  interference channels: The benefit of reconfigurable antennas,''
  \emph{Information Theory, IEEE Transactions on}, vol.~58, no.~6, pp.
  3766--3779, 2012.

\end{thebibliography}
\end{document}